\documentstyle[12pt]{article}
\begin{document}
\begin{center} \bf{CHIRAL QED IN TERMS OF CHIRAL BOSON WITH A GENERALIZED FADEEVIAN REGULARIZATION}\\
Anisur Rahaman, Durgapur Govt. College, Durgapur - !4, Burdwan, West Bengal, INDIA\\
e-mail: anisur@tnp.saha.ernet.in \end{center}

\vspace{.5cm}

\vspace{2cm}
\begin{center} \bf{Abstract} \end{center}
Chiral QED with a generalized Fadeevian regularization is considered. Imposing a chiral constraint a gauged version of Floranini-Jackiw
lagrangian is constructed. The imposition of the chiral constarint has spoiled the manifestly Lorentz 
covariance of the theory. The phase space structure for this theory has been determined. 
It is found 
that spectrum changes drastically but it is Lorentz invariant. Chiral fermion disappears from the spectra and the photon anquire mass as well.  Poincare algebra
has been calculated to show  physicial Lorentz invariance explicitely.

\newpage

Chiral boson in $(1 + 1)$ dimension  has been attracting the attention of the theoretical physicist for last few years [1 - 6]. 
Not only we find its
application in  quantum field 
theory but also it has wide application in the construction of string theory \cite{IMB, LAB}. The interaction of chiral boson 
with the gauhiral QED with a generalized Fadeevian regularization is considered. Imposing a chiral constraint a gauged version of Floranini-Jackiw
lagrangian is constructed. The imposition of the chiral constarint has spoiled the manifestly Lorentz
covariance of the theory. The phase space structure for this theory has been determined.
It is found
that spectrum changes drastically but it is Lorentz invariant. Chiral fermion disappears from the spectra and the photon anquire mass as well.  Poincare algebra
has
leads to consistent field theoretical models. So quantitative study of chiral boson and gauging of it have acquired a significant position in
$(1 + 1)$ dimensional quantum field theory.
Siegel proposed a manifestly
Lorent covariant action for chiral boson using auxiliary fields \cite{SIG}. The classical reparametrization symmetry of this action was not
maintained at the quantum mechanical level because of the gravitational anomaly. An alternative formulation was given by Floreanini
and Jackiw \cite{FJ} which had no minifestly Lorentz covariance to start with nevrtheless  physical Lorentz invariance was 
maintained there. Belucci,
Golterman and Petcher \cite{BEL} proposed a Lorentz covariant formulation which described the interaction of chiral boson with 
Abelian and 
non-Abelian gauge fields and discussed the Chiral Schwinger model as an example.

Chiral QED can be described either by the use of Dirac fermion as done by Jackiw and Rajaraman \cite{JR} or by the use of chiral(Weyl) 
fermion
as described by Harada \cite{KH}. Bosonization of the fermionic model in $(1+1)$ dimension needs regularization in order to remove
the divergence of the fermionic determinant. There are various way to regularize a theory and different type of counter term may appear
for different choices of regularization \cite{JR, KH, PM, AR, AR1},. Chiral Schwinger model has been found two be physicaly sensible 
for two different type
of masslike regularization \cite{JR, PM}. Certainly, one of these two is the regularization proposed by 
Jackiw and Rajaraman which removed the long
suffering of this model from nou-unitarity problem \cite{JR}. Another one is completely different in nature, commonly known as Fadeevian 
regularization in the literature, initially proposed by Mitra \cite{PM}. The former one was one parameter class of regularization 
whereas the regularization
proposed by Mitra was just a specific masslike terms for the gauge fields where no parameter was involved. Recently, Abreu {\it etal}.
\cite{WOT}
showed that a generalized (one parameter class of) Fadeevian regularization also exists which leads to physically sencible theory. 
Mitra's regulaizatio of course belonged to this class. The 
mass term proposed by Abreu {\it etal}. though contains two arbitrary parameter it ulitimately reducesd into a single parameter through
a constained rlation which appeard as a requirement to have Lorentz invariant spectra. Mitra initially used the Fadeevian regularization
to describe chiral QED where he used the chiral interaction of the Dirac fermion with the gauge field as Jackiw and Rajaraman did
in \cite{JR}. Mitra and ghosh described the chiral QED using chiral fermion \cite{MG}. It should be mentioned here that they did not 
derive it
from the fermionic lagrangian however it was the correct bosonised version of the Chiral QED in terms of chiral boson. The new
version of chiral QED with generalized fadeevian regularization is a description of chiral QED where chiral interaction of Dirac fermion
with the gauge field was considerd. So it would certainly be interesting to investigate whether this generalized Fadeevian regularization
does work for describing chiral QED in terms of chiral boson. To be more specific, whether a gauged version of Floreanini-Jackiw 
lagrangian is consitent with this  one parameter class fadeevian regularization. This motivates us to describe the theory of 
gauged chiral
boson with the generalized fadeevian regularization. 

The plan of this paper is the following. We impose a chiral constraint in the phase space of the theory proposed by 
Abreu {\it etal}. and obtain the effective gauged lagrangian of chiral QED described in terms of chiral boson. We then carry
out the hamiltonian formulation of this gauged lagrangian and find out the physical spectra. It is shown that inspite
of the absence of manifetly Lorentz covariance in the starting lagrangian the physical Lorentz invariance is maintained in 
the reduced space
of the theory.

Chiral Schwinger model is described by the following generating functional
\begin{equation}
Z[A] = \int d\psi d\bar\psi e^{\int d^2x{\cal L}_f},
\end{equation}
with
\begin{eqnarray}
{\cal L}_f &=& \bar\psi\gamma^\mu[i\partial_\mu + e\sqrt\pi A_\mu(1-\gamma_5)]\psi \nonumber \\
 &=& \bar\psi_R\gamma^\mu i\partial_\mu\psi_R +\bar\psi\_L \gamma^\mu(i\partial_\mu + 2e\sqrt\pi A_\mu)\psi_L 
.\end{eqnarray}
The right handed fermion remains uncoupled in this type of chiral interaction. So integration over this right handed part leads to 
field independent counter part which can be absorbed within the normalozation. Integration over left handed fermion leads to
\begin{equation}
Z[A] = exp{ie^2\over 2}\int d^2x A_\mu[M_{\mu\nu} - (\partial^\mu +\partial^\mu) {1\over \Box} (\partial^\nu + \partial^\nu)]A_\nu
.\end{equation}
$M_{\mu\nu} = ag_{\mu\nu}$, for Jackiw-Rajaraman regularization where the  constant $a$ represents the regularization ambiguity and
$M_{\mu\nu} = \left(\begin{array}{cc} 1 & \alpha \\
\alpha & \beta \\
\end{array}\right)\delta(x-y)$, for the generalized Fadeevian regularization. Here regularization ambiguity is represented by $\alpha$ and
$\beta $ ($\alpha$ and $\beta$ satisfy a constrained relation in order to render a Lorentz invariant spectrum). In this paper we are 
interested in the generalized Fadeevian regularization. We should mention here that the regularizatin
proposed by Mitra belongs to this class ,i.e., it is a special case of this general matrix $M_{\mu\nu}$, where $\alpha$ and $\beta$ takes
the value $-1$ and $-3$ respectively. 

This generating functional when written in terms of the auxiliary field $\phi(x)$ it turns out
to the following
\begin{equation}
Z[A] = \int d\phi e^{i\int d^2x {\cal L}_B},
\end{equation}
with
\begin{eqnarray}
{\cal L}_B &=& {1\over 2} (\partial_\mu\phi)(\partial^\mu\phi) + e(g^{\mu\nu} - \epsilon^{\mu\nu)}\partial_\nu\phi A_\nu + 
{1\over 2} e^2 A_\mu M^{\mu\nu}A_\nu \nonumber\\
&=& {1\over 2}(\dot\phi^2 - \phi'^2) + e(\dot\phi + \phi')(A_0 - A_1) + {1\over 2}e^2(A_0^2 + 2\alpha A_0A_1 + \beta  A_1^2)
.\end{eqnarray}
The momentum correxponding to the field $\phi$ is 
\begin{equation}
\pi_\phi = \dot\phi + e(A_0 - A_1) 
.\end{equation}
The hamiltonian obtained through the Legender transformation is
\begin{equation}
H_B = \int d^2x [\pi_\phi\dot\phi - {\cal L}], \end{equation}
where
\begin{equation}
{\cal H}_B = {1\over 2}[\phi_\phi - e(A_0 - A_1)]^2 + {1\over 2}\phi'^2 - 2e\phi'(A_0 - A_1) - 
{1\over 2}e^2(A_0^2 + 2\alpha A_0A_1 + \beta A_1^2).
\end{equation}

At this stage, we impose the chiral constraint 
\begin{equation}
\omega(x) = \pi_\phi(x) - \phi'(x) =  0, \end{equation}
which is a second class constraint itself since
\begin{equation}
[\omega(x), \omega(y)] = -2\delta'(x-y). \end{equation}
After imposing the constraint $\omega(x) =  0$ into the generating functional we have
\begin{eqnarray}
Z_{CH} &=& \int d\phi d\pi_\phi \delta(\pi_\phi - \phi') \sqrt{det[\omega, \omega]}e^{ i\int d^2x(\pi_\phi\dot\phi - {\cal H}_B)}
\nonumber \\
&=&\int d\phi e^{i\int d^2x{\cal L}_{CH}}
,\end{eqnarray}
with
\begin{equation}
{\cal L}_{CH} = \dot\phi\phi' -\phi'^2 + 2e(A_0 - A_1)\phi' + {1\over 2}e^2 [(\beta - 1)A_1^2 + 2(\alpha + 1)A_0A_1].
\end{equation} 
It is interesting to mention that we have obtained the gauged lagrangin for chiral boson from the bosonised Lagrangian with
generalized Fadeevian regularization just by imposing the chiral constraint in the phase space. Harada in \cite{KH}, obtaind the same type 
of result for the usual chiral scgwinger model with one parameter class of regularization proposed by Jackiw and Rajaraman. In the 
following section we will carry out the hamiltonian analysis of this gauged lagrangian adding the kinetic energy term for the gauge
field with the lagrangian density ${\cal L}_{CH}$.

So the starting lagrangian density in our case is
\begin{equation}
{\cal L} = {\cal L}_{CH} - {1\over 4} F_{\mu\nu}F^{\mu\nu}.
\end{equation}
Here $F_{\mu\nu}$ stands for the field strength for the electromagnetic field. From the standard definition the momenta corresponding
to the field $\pi_\phi$, $\pi_0$ and $\pi_1$ are obtained.
\begin{equation}
\pi_\phi = \phi',\label{MO1}
\end{equation}
\begin{equation}
\pi_0 = 0,\label{MO2}
\end{equation}
\begin{equation}
\pi_1 = \dot A_1 - A_0'.\label{MO3}
\end{equation}
Using the above equations it is straightforward to obtain the canonical hamiltonian through a lagender transformation which reads
\begin{equation}
H_C = \int dx[{1\over 2}\pi_1^2 + \pi_1A_0' + \phi'^2 - 2e(A_0 - A_1)\phi' + {1\over 2}e^2[(\beta -1)A_1^2 + 2(1 + \alpha)A_0A_1)].
\end{equation} 
Equation (\ref{MO1}) and (\ref{MO2}) are the primary constraints of the theory. Therefore, the effective hamiltonian is given by
\begin{equation}
H_{EFF} = H_C + u\pi_0 + v(\pi_\phi - \phi'),
\end{equation}
where $u$ and $v$ are two arbitrary lagrange multiplier. The constraints obtained in (\ref{MO1}) and (\ref{MO2}) have to be preserve
in order to have a consistent theory. The preservation of the constraint (\ref{MO2}), leads to new constraint which is the Gauss
law of the theory:
\begin{equation}
G = \pi_1' + 2e\phi' +  e^2(1 + \alpha)A_1 = 0. \label{GAUS}
\end{equation}
The consistency requirement of the constraint (\ref{MO1}) though does not give any new constraint  it fixes the velocity $v$ which
comes out to be
\begin{equation}
v = \phi' - e(A_0 - A_1). \label{VEL}
\end{equation}
The conservation of the Gauss law constraint, $\dot G = 0$, also gives a new constraint 
\begin{equation}
(1 + \alpha)\pi_1 + 2\alpha A_0' + (\beta + 1)A_1' = 0.\label{FINC}
\end{equation}
No new constraints comes out from the preservation of (\ref{FINC}). So we find that the phase space of the theory 
contains the following
four constraints. 
\begin{equation}
\omega_1 = \pi_0, \label{CON1}
\end{equation}
\begin{equation}
\omega_2 = \pi_1' + e\phi' +  e^2(1 + \alpha)A_1 = 0,\label{CON2}
\end{equation}
\begin{equation}
\omega_3 = (1 + \alpha)\pi_1 + 2\alpha A_0' + (\beta + 1)A_1' = 0,\label{CON3}
\end{equation}
\begin{equation}
\omega_4 = \pi_\phi - \phi'. \label{CON4}
\end{equation}
The four constraints (\ref{CON1}), (\ref{CON2}), (\ref{CON3}) and (\ref{CON4}) are all weak condition upto this stage. Treating this 
constraints as strong condition we obtain the following reduced hamiltonian.
\begin{equation}
H_R =  {1\over 2}\pi_1^2 + {1\over {4e_2}} \pi_1'^2 + {1\over 2}(\alpha - 1) \pi_1'A_1 + {1\over 4}e^2[(1 + \alpha)^2 - 2(1 - \beta)]A_1^2.
\label{RHAM}
\end{equation}
According to Dirac, Poission bracket gets invalidate for this reduced Hamiltonian \cite{DIR}. This reduced Hamiltonian however 
be consistent with the Dirac brackets which is defined by
\begin{equation}
[A(x), B(y)]^* = [A(x), B(y)] - \int[A(x) \omega_i(\eta)] C^{-1}_{ij}(\eta, z)[\omega_j(z), B(y)]d\eta dz, \label{DEFD}
\end{equation}
where $C^{-1}_{ij}(x,y)$ is defined by
\begin{equation}
\int C^{-1}_{ij}(x,z) [\omega_i(z), \omega_j(y)]dz = 1. \label{INV}
\end{equation}
For the theory under consideration
\noindent $C_{ij}(x,y) =$
\begin{equation}
 \pmatrix {0 & 0 & 2\alpha\delta'(x-y) & 0 \cr
0 & -2e^2(1+\alpha) \delta'(x-y) & e^2(1+\alpha)\delta(x-y) & 2e\delta'(x-y) \cr
& & -(\beta +1)\delta''(x-y)& \cr
2\alpha\delta'(x-y) & -e^2(1+\alpha)\delta(x-y)  & 2(\alpha+1)\times &  0 \cr
& +(\beta+1)\delta''(x-y) &(\beta+1)\delta'(x-y) & \cr 
0 & 2e\delta'(x-y) & 0 & 2\delta'(x-y) \cr}.
\label{MAT}
\end{equation}                

With the definition (\ref{DEFD}), and using equations (\ref{INV}) and (\ref{MAT}), we can compute the Dirac brackets
between the fields describing the reduced Hamiltonian $H_r$:
\begin{equation}
[A_1(x), A_1(y)]^* = {1\over {2e^2}}\delta'(x-y), \label{DR1}
\end{equation}
\begin{equation}
[A_1(x), \pi_1(y)]^* = {(\alpha -1)\over {2\alpha}}\delta(x-y),\label{DR2}
\end{equation}
\begin{equation}
[\pi_1(x), \pi_1(y)]^* = -{(1+\alpha)^2 \over {4\alpha}}e^2\epsilon(x-y).\label{DR3}
\end{equation}
Using the reduced hamiltonian (\ref{RHAM}), and the Dirac brackets (\ref{DR1}), (\ref{DR2}) and (\ref{DR3}), we obtain the 
following first order equations of motion
\begin{equation}
\dot A_1 = {(\alpha-1) \over {2\alpha}} \pi_1 + {1\over {2\alpha}} (2\alpha - \beta + 1)A_1', \label{EQM1}
\end{equation}
\begin{equation}
\dot \pi_1 = \pi_1' + {(\alpha-1)(\beta-1) \over {2\alpha}} e^2A_1. \label{EQM2}
\end{equation}
After a little algebra the equations (\ref{EQM1}) and (\ref{EQM2}) reduce to the following 
\begin{equation}
\partial_{-}\pi_1 = {(\alpha-1)(\beta-1) \over {2\alpha}} e^2A_1, \label{REQ1}
\end{equation}
\begin{equation}
\partial_{+}A_1 = {(\alpha-1)\over {2\alpha}} \pi_1 + {1\over {2\alpha}}(4\alpha - \beta + 1)A_1' .\label{REQ2}
\end{equation}
The above two equations (\ref{REQ1}) and (\ref{REQ2}) satisfy the following Klein-Gordon Equation
\begin{equation}
(\Box - {(\alpha-1)^2\over {\alpha}})\pi_1 = 0\label{SPEC},
\end{equation}
provided $\alpha$ and $\beta$ satisfy the constrained relation
\begin{equation}
4\alpha - \beta + 1 = 0. \label{RES}
\end{equation}

The equation (\ref{SPEC}), represents a massive boson with square of the mass is given by
$m^2 = {{-(1-\alpha)^2}\over \alpha
}$. $\alpha$ must be negative for the mass of the boson to be physical. Unlike the Abreu {\it etal}., there is no massless degrees of 
freedom in this situation. Ofcourse,
the constraint
structure is different here and therefore, reduction of massless degrees of freedom is not unnatural. 
If we look into the Mitra and Ghosh's
 description 
we find that we can land into their results for the specific value of the parameter $\alpha= -1$ and $\beta = -3$. So what we achieved from
this model is what follows. The spectrum contains  only a massive boson. Unlike the Mitra and Ghosh's  description, it is 
parameter dependent. One can 
think of it as the photon acquires mass and the fermion gets confined, i.e., completely eaten up during the process. In this situation
it is the chiral fermion the bosonised version of which is a chiral boson.
We have noticed that for the specific value $\alpha = -1$ and $\beta = -3$, the
model maps into Mitra and Ghosh's model. Hence the present one is the  generic situation. The restriction 
(\ref{RES}), is also satisfied by this specified value of $\alpha$ and $\beta$. This may be used as a check for
consistancy.

We have found that the gauged lagrangian for chiral boson obtained here  does not have Lorentz covariance,
howeverit gives Lorentz invariant spectrum.
So our next task is to check whether Poincare algebra is satisfied in the reduced phase space or not. A closer look reveals that 
the invariance is not maintained
in the whole subspace of the theory but in the physical subspace it is maintained inspite of having such a deceptive appearence. To show 
the 
Lorentz invariance we
have to calculate the Poincare algebra. There are three elements in this algebra, the hamiltonian $H_{R}$, the momentum $P_{R}$ and the boost 
generator $M_{R}$ and they have to stisfy the following relation in $(1 + 1)$.

\begin{equation}
[P_{r}(x), H_{R}(y)]^* = 0, 
[M_{R}(x), P_{R}(y)]^* = -H_{R},
[M_{R}(x), H_{R}(y)]^* = -P_{R}.\label{POIN}
\end{equation}

Hamiltonian has already been given in (\ref{RHAM}) and the momentum density reads  
\begin{eqnarray}
{\cal P}_R &=& \pi_1A_1' + \pi_\phi\phi', \nonumber \\
         &=& {1\over {4e^2}}\pi_1^2 + {1\over 2}(1-\alpha)\pi_1A_1' + {1\over 4}e^2(1+\alpha)^2A_1^2
\end{eqnarray} 
The Boost generator written in terms of hamiltonian density  and momentum is
\begin{equation}
M_{R} = tP_{R} + \int dx x {\cal H}_{R} \label{MOM} \end{equation}
Straightforward calculations shows that equation(\ref{POIN}) is satisfied provided the restriction (\ref{RES}), on $\alpha$ nad $\beta$ 
is maintained. This restriction also appeared in the process of finding out the Lorentz 
invariant  theoretical spectrum (\ref{SPEC}) from the
equations of motion. The above calculations therefore confirms the physical Lorentz invariance of this model. This explicit Lorentz 
invarinance suggests that there may be a Lorentz covariant structure for this theory.

In this paper we have formulated the gauged version of Floreanini-Jackiw Lagrangian just by imposing a chiral constraint into the 
action
of the chiral Schwinger model with a generaliged Fadeevian regularization proposed by Abreu {\it etal}. The effective action obtained after 
imposing the chiral constarint acquires a deceptive appearence. To be specific there were no term which had manifestly Lorentz 
covariance. However physical Lorentz invariance is found to maintained. 
After studing the phase space structure we find that the constraint structure are completely different from the description Abreu {\it etal}.
and  consequntly, a drastic change in the spectra is noticed. The chiral boson which can be thought of in terms of chiral 
fermion disappears. Photon acquires masss as well. The mass of the photon is found to be identical to the mass of the masssive boson as 
obtained by Abreu {\it etal}. This new description of chiral QED in terms of 
chiral boson with the generaliged Fadeevian regularization reminds us the results of the usual Scgwinger model where also photon acquire 
mass and fermions 
get confined. The fermion confined here is chiral (Weyl) type whereas the fermion confined in the Schwinger model was Dirac type because
the lagrangian with which we started contains chiral boson which in $(1 + 1)$ dimension is equivalent to chiral fermion. Thus Schwinger
mechanism works for chiral Schwinger model too for this generalized Fadeevian regularization.

In the study of $(1 + 1)$ dimensional QED and chiral QED it has been found several times that regularization has a  crucial link 
with the confinement \cite{JR, PM, AR, AR1, SCH}. Though 
it is fair to say that till now there is no such qualitative study which shows a direct link between these two. This analysis also
shows the repeatation of the same fact, i.e., we notice a change in the confinement scenario with the variation of 
regularization. There is a natural thrust for finding out a direct link between confinement and regularization. This type of study may throw some
light on this issue.
More qualitative invesigation ofcourse be needed.

Another interesting aspect of this theory is the the result we obtained in connection with the Poincare invariance property of its 
lagrangian. We find that only in the physical subspace the Poincare invariance is maintained.
To ensure the Poincare invariance we generally start with a Lorentz covariant theory. But hre the situation is not 
like that. To start with it had no Lorentz covariance
but the constraint structure of this theory is so mysterious that it ultimately reduced to a theory which satisfy correct
Poincare algebra. This type of study may help the general investigation of finding how constraint structure dictates or threatens 
Poincare symmetry. More qualitative study is also needed in this issue too.

I would like to acknpwledge Prof. P. Mitra of Saha Institute of Nuclear Physics, Calcutta, for helpfull discussion.

\end{document}